\documentclass[12pt]{article}
\usepackage{epsfig}
\begin{document}

\begin {center}
{\Large \bf {An alternative interpretation of Belle data on 
$\gamma \gamma \to \eta '\pi ^+\pi ^-$}}
\vskip 5mm
D.V.Bugg
\bigskip 
{\it Queen Mary, University of London, London E1\,4NS, UK}.
\end {center}

\begin{abstract}
Belle data on $\gamma \gamma \to \eta' \pi \pi$ are refitted
using a broad $J^{PC}=0^{-+}$ peaking in the mass range 2250-2300 and 
$X(1835)$, but without $\eta (1760)$.  
There is the possibility that the broad $0^{-+}$ signal may be identified
with the $0^{-+}$ glueball predicted originally by Morningstar and Peardon.
The $X(1835)$ is confirmed to have a resonant phase variation.  

\end {abstract}

\section {Introduction}

The Belle collaboration presents new data on 
$\gamma \gamma \to \eta '\pi ^+ \pi ^-$ \cite {Belle}.
In their Fig. 3, there is evidence for fine structure in the 
$\eta '\pi \pi$ mass range 1700--1900 MeV.
On a larger scale, there is a conspicuous broad peak centred at $2300$ MeV
with a full width at half-maximum of $\sim 750$ MeV, see their Fig. 2(b).
As a shorthand, this peak will be called $Y(2300)$.
No fine structure is visible in this broad peak from 2000 to 2800 MeV.
It deserves attention, since it could be the $0^-$ glueball 
predicted by Morningstar and Peardon  near this mass \cite {Morningstar}.

It would not be surprising if this glueball is very wide.
Zou, Dong and I have drawn attention to a very broad $J^{PC}=0^-$ signal 
observed in $J/\psi$ radiative decays \cite {Zou}.
It accounts in a simple way for successive peaks in $J/\psi \to \gamma X$, 
where $X \to \rho \rho$, $\omega \omega$, $K^*\bar K^*$ and $\phi \phi$ 
channels, with flavour-blind coupling strengths.
$J/\psi$ radiative decays are dominated by  $\gamma GG$, where $G$ are 
gluons.
The conclusion of Ref. \cite {Zou} was that there is a broad $0^{-+}$ 
signal consistent with a glueball with mass $M = 2190 \pm 50$ MeV and width 
$\Gamma = 650 \pm 100$ MeV. 
The half-width of the lower side of the peak in Belle data is 300--350
MeV.
It would be a pity to miss the $0^-$ glueball if it is really there. 

Belle base their analysis on the claim by DM2 to observe $\eta (1760)$ in
data on $J/\psi \to \gamma (\pi ^+\pi ^-\pi ^+\pi ^-)$ \cite {DM2}.
However, an analysis of Mark III data on the same channel showed that 
the 1760 MeV peak has $J^{PC}=0^{++}$, though it does sit on a large, 
broad $0^-$ background \cite {Mark3}.
[A technical detail is that there is no interference between
$0^{++}$ and $0^{-+}$ after summing over spin orientations of the $J/\psi$.]
Peaks at 1500 and 2105 MeV were also fitted with $J^{P}=0^{+}$.
Furthermore, high statistics data of the Fermilab E760 experiment
\cite {E760} on $\bar pp \to (\eta \eta)\pi ^0$ fit all three peaks 
accurately with the same mass and width for these $J^P=0^+$ states; 
$J^P=0^-$ is forbidden in $\eta\eta$ by the Pauli principle.
Many authors have been confused by the fact that the PDG \cite {PDG} does 
not mention Ref. \cite {DM2} under $\eta (1760)$, though it is listed under
$f_0(2100)$. 

The existence of $\eta (1760)$ therefore rests on 
(a) the BES I analysis \cite {BES1}, where $M=1760 \pm 35$ MeV, 
$\Gamma \sim 250$ MeV, (but quoted as not well determined),
(b) the BES II analysis of $J/\psi \to \gamma (\omega \omega)$, 
$M = 1744 \pm 10(stat) \pm 15(syst)$ MeV, $\Gamma = 244 ^{+24} _{-21} \pm 25$
MeV \cite {BES2}. 
There is a serious objection to this second source.
A well known relation, coming from SU(2) symmetry, is that an isospin $I=0$
resonance should have equal couplings to $\omega \omega$ and $\rho^0\rho^0$,
because light quarks do not discriminate between charges. 
There are three charge states for $\rho \rho$, so the relation is
normally written $g^2(\rho \rho) = 3g^2(\omega \omega)$, where $g$ are
coupling constants.
This relation applies equally well to $q\bar q$ states, hybrids and 
glueballs, which all obey SU(2).

The branching fraction quoted for production of $\eta (1760)$ in the BES II
$\gamma \omega \omega$ data is $(1.98 \pm 0.08(stat) \pm 0.32(syst)) \times
10^{-3}$.
This is larger than the $\rho^0 \rho^0$ weighted mean branching fraction 
from DM2 and Mark III over the entire mass range up to 2 GeV, namely 
$\sim (1.23 \pm 0.25) \times 10^{-3}$.
It should lead to a huge $\rho \rho$ peak at 1760 MeV, in  
disagreement with BES I, DM2 and Mark III data.
The DM2 collaboration did claim a small $\eta (1760)$ signal in data on
$\rho \rho \to \pi ^+\pi ^-\pi ^+\pi ^-$, but without observing any phase
variation. 
Their branching fraction was a factor 4 smaller than BES II claim in 
$\gamma \omega \omega$.
The $\pi^+\pi^-\pi +\pi ^-$ data are experimentally much cleaner than
$\gamma \omega \omega$, $\omega \to \pi ^+\pi ^-\pi ^0$, where there are
5 photons, hence large combinatoric problems. 

\begin{figure}[htb]
\begin{center}
\vskip -5mm
\epsfig{file=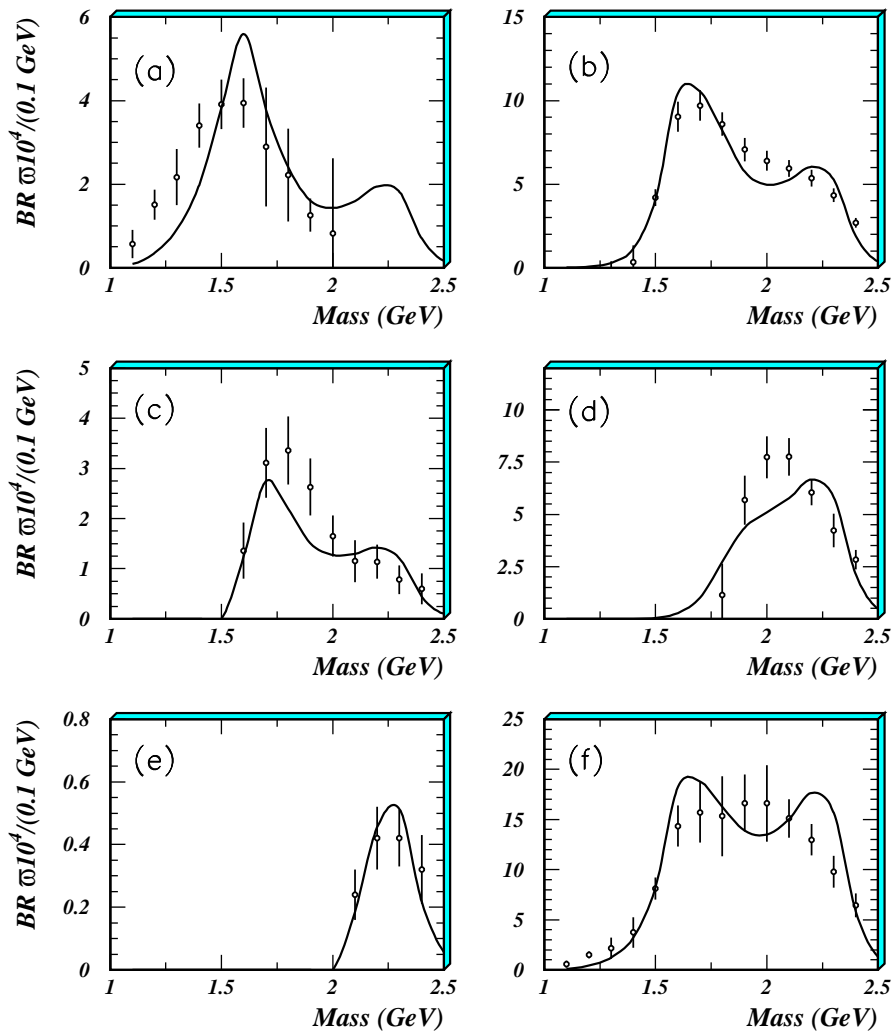,width=12cm}
\vskip -8mm
\caption {Data and fits to $J/\psi \to $ (a) $\eta \pi \pi$, (b) $\rho \rho$,
(c) $\omega \omega$, (d) $K^*\bar K^*$, (e) $\phi \phi$ and (f) all channels,
from Ref. \cite {Zou}.
Points with errors show averages of data from BES I, DM2 and Mark III.
Reproduced with permission.
}
\end{center}
\end{figure}
Fig. 1 reproduces the magnitudes of branching fractions used in Ref. 
\cite {Zou} for five channels and the total.
This analysis used fully analytic amplitudes where the denominator of
the amplitude takes the form
\begin {eqnarray}
D(s) &=& M^2 - s - m(s) - i\sum _j g^2_j \rho_j(s) \\
m(s) &=& \frac {s-M^2}{\pi}\, P \int ^\infty _{sthr} 
\frac {\sum _j g^2_j\rho_j(s')\, ds'}{(s' - s)(s' - M^2)}.
\end {eqnarray}
Here $P$ stands for the Principal Value integral;
$sthr$ is the threshold for each channel $j$.
Note that this is not an `optional extra'; it is a requirement of 
analyticity. 
Wherever an opening channel produces a peak, there is a cusp, which need
not be resonant.

The $\rho \rho $ channel shown in panel (b) peaks just below 1600 MeV 
because of the $L=1$ centrifugal barrier for production.
In panels (b) and (c), amplitudes are restricted to obey the SU(2) relation.
The peak in $\omega \omega$ can be confused with a resonance,
but at half-height it has a width of $\sim 350$ MeV.
In Ref. \cite {Zou} it was shown that there was no pole in the 
$\omega \omega$ amplitude, nor in $\eta\pi \pi$ in this mass range.

Section 2 discusses first a fit to the broad peak considering only
$J^P = 0^- \to \eta \sigma$.
The objective is to refit the Belle data without $\eta (1760)$.
The $X(1835)$ component is needed to fit the mass distribution. 
Note that the $[\eta \sigma]_{L=1}$ decay (where $L$ is orbital
angular momentum) is forbidden for $\gamma \gamma \to 1^{++}$ 
(Yang's theorem). 

However, this is not the whole story.
Belle's Fig. 6 displays the $\pi ^+\pi ^-$ mass distribution for
$\eta '\pi \pi$ events in the mass range 2200--2700 MeV/c$^2$.
There is an $f_2(1270)$ peak in $\eta \pi \pi$, with an intensity 
$\sim 85\%$ of that of the $\sigma$ near 1 GeV.
A likely amplitude producing the $f_2$ is $Y(2300) \to [\eta 'f_2]_{L=2}$,
with the result that there is a single broad $0^-$ resonance.
This possibility is investigated in sub-section 2.1. 

\section {Initial fit to the $\eta '\pi^+\pi^-$ peak}
Ideally, one would adopt the same approach as used in Ref. \cite {Zou},
where a coupled channel fit to many open channels was made.
However, presently there is limited knowledge of some important 
amplitudes in the mass range above 1800 MeV, e.g. 
$\gamma \gamma \to K\bar K\pi$ and $\eta \pi \pi$. 
Only the simplest parametrisation for the broad
component visible in Belle data can be used at the moment, but it is
essential to accomodate the opening of $\eta'\sigma$ and $\eta'f_2$
phase space $\rho$.
The simplest Breit-Wigner amplitude $f(s)$ with these features is
\begin {eqnarray}
f(s) &=& M\Gamma (s)/[M^2 - s - iM\Gamma_0] \\
\Gamma (s) &=& FF(k)g^2(\eta '\pi\pi)\rho (\eta'\pi \pi).
\end {eqnarray}
If other decay modes of $Y(2300)$ exist, such as $K\bar K\pi$,
$\eta \pi \pi$, $\omega \omega$ and $\rho \rho$, then $\Gamma_0$
will be summed over all decay modes.
Here it is taken as a constant, the simplest possibility. 
For the numerator, a form factor $FF(k) = \exp (-\alpha k^2)$ is 
used, where $k$ is the momentum of the $\eta '$ in the overall centre 
of mass.
Good fits are obtained for $\alpha = 1.5$ to 3.0 (GeV/c)$^{-2}$,
consistent with data on $J/\psi$ radiative decays \cite {Zou}. 
The form factor arises from convolution of form factors for
the outgoing $\eta \pi \pi$ final state and the initial $\gamma \gamma$ 
interaction. 
The phase space factor $\rho$ is obtained from the integral over 3-body 
phase space, given by equations (39.19) and (39.20) of the Particle Data 
Book \cite {PDG}.
For fits where the $0^-$ initial state decays to both $\eta '\sigma$
and $\eta' f_2$, followed by decays of both channels to $\pi ^+\pi ^-$,
$\Gamma (s)$ of Eq. (3) needs to include fully coherent interferences 
between both channels.
The $\sigma$ is parametrised by the amplitude given in Ref. 
\cite {sigpole}, Table 1, entry (iii); this parametrisation allows not only 
for decays $\sigma \equiv f_0(500) \to \pi \pi$, but also to $K\bar K$ 
(which is quite significant), $\eta \eta$ (small) and $4\pi$ (large above 
1350 MeV, but affecting only $\eta '\pi \pi$ masses above $\sim 2250$ MeV
and rather uncertain in magnitude). 
The $X(1835)$ is included in the fit multiplied 
by an isobar model phase factor $\exp (i\delta)$, with constant $\delta$.

\begin{figure}[htb]
\begin{center}
\epsfig{file=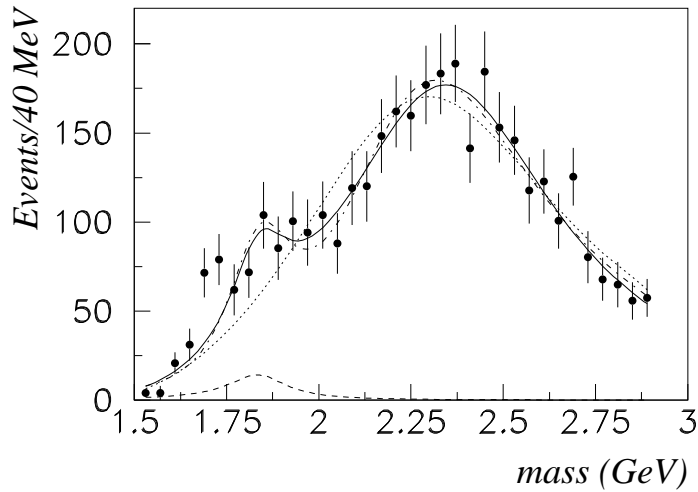,width=10cm}
\vskip -7.5mm
\caption {Fits to Belle data for 
$\gamma \gamma \to \eta' \pi ^+\pi ^-$ from their Fig. 2(b); full line: 
fit with $Y(2300)$ and $X(1835)$; dashed curve, the fitted intensity 
of $X(1835)$; dotted curve, fit with only $Y(2300)$; 
chain curve, the full intensity including $[\eta ' f_2]_{L=2}$.}
\end {center}
\end{figure}
Three curves on Fig. 2 show the fit to Belle data, initially using only 
$0^- \to \eta \sigma$ over the whole mass range up to 2800 MeV; 
a fourth curve shows the effect of including $0^- \to [\eta ' f_2]_{L=2}$ 
as described in sub-section 2.2. 
The full line shows the fit including $X(1835)$. 
Its mass and width are allowed to vary within the statistical and 
systematic errors quoted by BES III \cite {BES3}.
A resonant phase variation is required for $X(1835)$ to account for
interference with $Y(2300)$.
The mean $\chi^2$ per point is 1.25 after allowing for the 7 fitted
parameters.
Two points at 1.69 and 1.73 lie 2.8 and 2.4 standard deviations above the 
fit, but would require a very narrow peak inconsistent with the broad 
$\omega \omega$ cusp.
The dotted curve shows the fit without $X(1835)$.
The contribution from $X(1835)$ to the full line is an 8.2 standard
deviation effect, after correcting deviations from the fit
by dividing by 1.25 and allowing for the change in the number of
fitting parameters from 7 to 4.
So the $X(1835)$ is confirmed as a $0^{-+}$ resonance.

Using the whole mass range, the mass fitted to the 
broad $0^-$ signal in Fig. 2(a) is  $2300 ^{+75}_{-80}$ MeV and 
$\Gamma = 750 ^{-40}_{+45}$ MeV, where the signs display the correlation 
between $M$ and $\Gamma$.
The dashed curve at the bottom of Fig. 2 shows the optimum contribution 
from $X(1835)$.
The $\eta (2300)$ intensity without $X(1835)$ is shown by the dotted 
curve.  

If $\gamma \gamma$ couples to $\omega \omega$, as seems likely,
the full width of the cusp due to this threshold is $\sim 350$ MeV at
half-maximum from Fig. 1(c).
This will alter the entire $\eta(2300)$ mass distribution slightly,
but cannot be predicted without the $\gamma \gamma \to \omega \omega$ 
coupling constant.

\subsection {The effect of $\eta ' f_2(1270)$}
The $J^P = 0^- \to [\eta ' f_2]_{L=2}$ contribution provides a ready
explanation of $f_2$ production.  
Their coherent sum is deduced using the relative magnitudes of $f_2$
and $\sigma$ signals in Fig. 5 of Belle. 
The phase space factor for $f_2$ includes a standard $L=2$ centrifugal 
barrier factor $B(s)= 9k^4/(9 + 3k^2R^2 + k^4R^4)$.
The value $R=0.725$ fm is used.  
This assumes the same radius of interaction as the Gaussian 
form factor $FF=\exp -[k(GeV/c)/\hbar c]^2R^2/6 = \exp (-\alpha k^2)$,
using the optimum  $\alpha = 2.25$ (GeV/c)$^{-2}$.
Following the isobar model approach, the amplitude requires a factor 
$\exp {i\delta}$, where $\delta$ is a constant, $25^\circ$.
The angular distribution between $\eta '$ and $f_2$ is isotropic,
like $\eta '\sigma$.
A summation is made over the $f_2$ decay.

The $[\eta ' f_2]_{L=2}$ amplitude produces a slow rise centred at
2400 MeV.
In order to accomodate this rise, the width fitted to $Y(2300)$ 
decreases by $\sim 100$ MeV to $650$ MeV,
with an associated reduction in the mass of $\sim 50$ MeV.
These values provide an estimate of systematic errors for the mass and 
width of $Y(2300)$.
This extra component allows slightly more freedom in the low mass 
range considered in the previous section.
The significance of the $X(1835)$ contribution falls from 8.2 to
7.1 standard deviations, but there is no essential change to the
fit, which is shown by the chain curve in Fig. 2. 
It produces (i) a slightly larger $X(1835)$ peak, (ii) a $Y(2300)$ mass
reduced to $\sim 2250$ MeV, and (iii) a  higher tail near
2700 MeV caused by the rising centrifugal barrier.  
It does produce an angular distribution resembling Belle's Fig. 5.
However, a full fit to the Dalitz plot will be required to be precise
about the angle and energy dependence of this term, if indeed it is
present.

Belle suggest a contribution to the $Y(2300)$ peak with $J^P = 2^+$.
This would produce the final state ${\eta 'f_2(1270)}_{L=1}$.
There is also the possibility of $J^P = 2^- \to [\eta 'f_2]_{L=0}$.
Only an analysis of the Dalitz plot can identify such contributions.
A remark is that the angular momentum analysis of these cases needs 
to obey gauge invariance for the photons.
In their centre of mass, they have only helicity amplitudes
$|1,1>$ and $|1,-1>$, but no $|1,0>$ component.
If axes are used in the $\gamma \gamma$ rest frame with the $z$-axis
along the direction of the photons, only the $x$ and $y$ components
of the spin 2 combination contribute; their intensities add incoherently.
To describe the $f_2$ it is necessary to rotate axes to its direction
and then make a Lorentz boost to its rest frame.
Using rotation matrices and Clebsch-Gordan coefficients is prone to
mistakes, but possible.
A simpler approach is the so-called method of Wick rotations.
Axes are first rotated to the direction of the $f_2$ using angles
$\theta, \phi$.
After expressing  momenta of pions from the $f_2$ decay in this system, 
the Lorentz transformation of the pions is made to the $f_2$
rest frame. 
Finally, the axes in that frame are rotated back through angles 
$-\theta,-\phi$, taking care that the product of the two rotations is 
the $3 \times 3$ matrix with unit diagonal elements.
The two rotation matrices cancel.
The Lorentz boost changes the angles of the pions from the $f_2$ decay
between the $\gamma \gamma$ rest frame and the $f_2$ rest frame,
but the usual Clebsch-Gordan coefficients and spherical harmonics 
give the correct decay amplitude, which also requires the usual
centrifugal barrier factors.
A check on the procedure is that all amplitudes, including those
for $\eta ' \pi \pi$ used above, are orthogonal when integrated over
all angles.
If the experimental acceptance is included using the Monte Carlo
simulation, the effects of cuts and acceptance are immediately apparent 
in the interferences between amplitudes.
The third method is to use covariant tensor expressions for amplitudes, 
but this is unfamiliar to most experimentalists.
 
\section {Concluding Remarks}
The glueball hypothesis is clearly a matter of conjecture, but
is worth following up with further studies of 
$\gamma \gamma \to \eta\pi\pi$ and $K\bar K\pi$ in particular.
These  are important for checking whether the $Y(2300)$ decays 
flavour blind.
The observed broad, well defined peak in $\eta' \pi \pi$ does not
look like a conventional $q\bar q$ state, where masses are typically 
below 300 MeV.
The lower side of this peak requires interference with $X(1835)$
with $J^{PC}=0^{-+}$ in agreement with BES III; the $X(1835)$ has a 
definite resonant phase variation. 

At present, there is no clear benchmark for the mass scale of the
$0^-$ glueball, but Lattice estimates are in the mass range 2250 
\cite {Meyer} to 2590 MeV \cite {Morningstar}.
Mass ratios are better predicted than absolute values.
Morningstar and Peardon predict that it will
have a mass $1.50 \pm 0.04$ times that of the 
$0^+$ glueball and $1.08 \pm 0.04$ times that of the $2^{++}$ glueball.
A second $0^+$ glueball is also predicted in this mass range. 
It is possible that the $2^+$ glueball is to be identified with the
$f_2(1950)$, which is observed in decays to $\eta \eta$, $4\pi$, $KK$ and
$KK\pi\pi$; it has a width of $472 \pm 18$ MeV \cite {PDG}.
There is a new quenched Lattice Gauge calculation of the glueball spectrum 
by Gregory et al. \cite {Gregory}, who give a careful review.
The present theoretical situation is that eigenvalues, i.e. masses, 
have been calculated.
These are based on couplings to gluons.
What is presently not clear is the effect of decays to $q\bar q$ nonets.
These are in most cases still buried in two point correlation functions
which include glueball and $q\bar q$ combinations \cite {Steele}. 

In BES III data for the $\eta '$ channel, there are further peaks at 2122 
and 2376 MeV with widths of $83 $ MeV for both. 
A natural interpretation of the $X(1835)$ is that it is the $n=3$ radial
excitation of $\eta (958)$ $(n=1)$ and $\eta(1405/1475)$ $(n=2)$.
For the latter, the evidence for two separate states is not conclusive; 
the stronger $\eta (1475)$ decays dominantly to $KK^*(890)$ in a 
P-wave, and the $k^3$ increase of the P-wave shifts the average mass of 
the peak up by $\sim 35$ MeV.
Also Wu et al. \cite {Wu} have proposed  an interpretation of 
$\eta (1405)$ in terms of a triangle graph where $KK$ from $KK^*$ decay 
rescatter to $\eta \pi$. 
In  $J/\psi \to \gamma (\eta \pi \pi)$, there is a large dispersive peak 
in Fig. 1(a) at 1500 MeV producing a strong enhancement of 
$\eta(1405) \to \eta\pi\pi$.
Achasov and Shestakov suggested in 1985 a natural explanation for a
very broad $J^{PC}=0^-$ signal observed in $J/\psi$ radiative decays to
$\rho \rho$ and $\omega \omega$ \cite {achasov1}.
In a later paper they suggested why $\eta (1440)$ and $\eta (1475)$
are not observed in $\gamma \gamma$ collisions \cite {achasov2}; further 
consequences of this suggestion were studied there.
A recent third paper reviews the question comprehensively including the
latest data and makes recommendations for further work \cite {achasov3}. 

It is not yet established that the peaks at 2122 and 2376 MeV have
$J^P=0^-$. 
However, if that is the case, mixing with the broad
$Y(2300)$  would enhance their visibility.
The gluon interaction is likely to be of short range, judging by the
funnel potential, but mixing with $q\bar q$ components which peak
at larger radii reduces the zero-point energy.
The sequence of peaks from $\eta (958)$ to
2376 MeV lies close to a straight trajectory of $M^2$ v $n$ with
a slope of 1.18 GeV$^2$, like that observed in Crystal Barrel
data for many resonances, namely $1.143 \pm 0.013$ GeV$^2$ \cite {survey}.
 
In summary, the Belle data can be fitted well with just $X(1835)$ and
a broad $J^{PC}=0^{-+}$ signal.
An $f_2$ component arises naturally from $0^- \to [\eta 'f_2]_{L=2}$
but that hypothesis needs confirmation from a full analysis of the Dalitz 
plot.

\end {document}